\documentclass[namedreferences]{SolarPhysics}
\usepackage[optionalrh,natbib]{spr-sola-addons} % For Solar Physics
\usepackage{graphicx}                    % For eps figures, newer & more powerfull
\usepackage{amssymb}                    % useful mathematical symbols
\usepackage{color}                       % For color text: \color command
\usepackage{url}                         % For breaking URLs easily trough lines
\newcommand{\adv}{    {\it Adv. Space Res.}}

\newcommand{\aap}{    {\it Astron. Astrophys.}}

\newcommand{\apj}{    {\it Astrophys. J.}}
\newcommand{\apjl}{    {\it Astrophys. J. Lett.}}

\newcommand{\grl}{    {\it Geophys. Res. Lett.}}

\newcommand{\jgr}{    {\it J. Geophys. Res.}}

\newcommand{\solphys}{{\it Solar Phys.}}

\newcommand{\ssr}{    {\it Space Sci. Rev.}}

\begin{document}

\begin{article}

\begin{opening}

\title{Constraining the Kinematics of Coronal Mass Ejections in the Inner Heliosphere with \textit{In-Situ} Signatures}

%%%%%%%%%%%%%%%%%%%%%%%%%%%%%%%%%%%%%%%%%%%%%%%%%%%
%% Authors Names
%
\author{T.~\surname{Rollett}$^{1,2}$\sep
        C.~\surname{M\"ostl}$^{1,2}$\sep
        M.~\surname{Temmer}$^{1,2}$\sep      
        A.M.~\surname{Veronig}$^{1}$\sep
        C.J.~\surname{Farrugia}$^{3}$\sep
        H.K.~\surname{Biernat}$^{1,2}$\sep}

%%%%%%%%%%%%%%%%%%%%%%%%%%%%%%%%%%%%%%%%%%%%%%%%%%%
%% Runningheads
%
\runningauthor{T.~Rollett \textit{et al.}}
\runningtitle{}

%%%%%%%%%%%%%%%%%%%%%%%%%%%%%%%%%%%%%%%%%%%%%%%%%%%
%% Affilations 
%
\institute{$^{1}$ Institute of Physics, University of Graz, A-8010, Austria
                  email: \url{tanja.rollett@uni-graz.at}\\ 
           $^{2}$ Space Research Institute, Austrian Academy of Sciences, Graz A-8042, Austria\\
           $^{3}$ Space Science Center and Department of Physics, University of New Hampshire, Durham, New Hampshire, USA\\}

%%%%%%%%%%%%%%%%%%%%%%%%%%%%%%%%%%%%%%%%%%%%%%%%%%%
%%% Abstract 
\begin{abstract}
We present a new approach to combine remote observations and \textit{in-situ} data by STEREO/HI and \textit{Wind}, respectively, to derive the kinematics and propagation directions of interplanetary coronal mass ejections (ICMEs). We use two methods, Fixed-$\phi$ (F$\phi$) and Harmonic Mean (HM), to convert ICME elongations into distance, and constrain the ICME direction such that the ICME distance-time and velocity-time profiles are most consistent with \textit{in-situ} measurements of the arrival time and velocity.
The derived velocity-time functions from the Sun to 1~AU for the three events under study (1\,--\,6 June 2008, 13\,--\,18 February 2009, 3\,--\,5 April 2010) do not show strong differences for the two extreme geometrical assumptions of a wide ICME with a circular front (HM) or an ICME of small spatial extent in the ecliptic (F$\phi$). Due to the geometrical assumptions, HM delivers the propagation direction further away from the observing spacecraft with a mean difference of $\approx 25^\circ$.
\end{abstract}

%%%%%%%%%%%%%%%%%%%%%%%%%%%%%%%%%%%%%%%%%%%%%%%%%%%
%% Keywords
%
%\keywords{}

\end{opening}
%-------------------------------------------------

%%%%%%%%%%%%%%%%%%%%%%%%%%%%%%%%%%%%%%%%%%%%%%%%%%%
%% Sections
%
\section{Introduction}
\label{s:intro}

Coronal mass ejections (CMEs) are manifestations of the most powerful eruptions on the Sun and are expulsions of a huge amount of plasma and the embedded magnetic field. Their velocities range between a few 100 km~s\textsuperscript{$-1$} up to more than 3000 km~s\textsuperscript{$-1$}. The frequency of occurrence correlates with the solar cycle, and faster and more powerful events are more common during the solar maximum phase. They can have wide latitudinal and longitudinal extents and their typical masses are of the order of $\approx10^{12} - 10^{13}$ kg \citep[\textit{cf.}][]{gos74,hun97}. CMEs propagating from the Sun through the interplanetary space are called interplanetary CMEs (ICMEs).

Before 2006 it was not possible to directly link CMEs and their properties as measured \textit{in-situ}. A milestone in the investigation of CMEs is the NASA \textit{Solar TErrestrial RElations Observatory} \citep[STEREO:][]{kai08} mission with its twin satellites STEREO-AHEAD (A) and STEREO-BEHIND (B). STEREO-A is leading the Earth in its orbit around the Sun and STEREO-B is following. The \textit{Heliospheric Imagers} \citep[HI1 and HI2:][]{eyl09} are part of the \textit{Sun-Earth Connection Coronal and Heliospheric Investigation} instrument suite \citep[SECCHI:][]{how08,har09} onboard the two STEREO spacecraft and they enable us for the first time to observe solar transient events from two vantage points from outside the Sun-Earth line. Such observations constitute a unique way of investigating the behavior of ICMEs all the way from the Sun to Earth. In addition, the possibility to study solar minimum events is a big advantage because it is necessary to do case studies of CMEs showing simple and well-defined remote as well as \textit{in-situ} signatures.

There are different methods that can be used to infer the direction of propagation of an ICME. Some methods use single spacecraft observations, \textit{e.g.}\ the Fixed-$\phi$ fitting method \citep{she99, rou08} or the Harmonic Mean fitting procedure \citep[][]{lug10,moe11} while others use data from both spacecraft, \textit{e.g.}\ the triangulation method by \cite{liu10a}, extended to circular fronts by \cite{lug10b}.

The aim of this work is to analyze the kinematics and propagation directions of a set of CMEs up to a distance of 1~AU. For this we developed a method based on a combination of remote and \textit{in-situ} measurements. We constrain measurements from time-elongation plots (Jmaps), produced out of the white-light HI1/HI2 images along the ecliptic plane, with the \textit{in-situ} measured arrival time and arrival velocity. The best matches deliver propagation directions and velocity profiles based on the geometric assumptions we use, for either very wide or narrow ICME fronts. This should serve as a basis to develop adequate methods to predict the arrival times of Earth-directed CMEs.

\section{Data}
\label{s:data}

The \textit{Heliospheric Imagers} \citep[HI:][]{eyl09} onboard STEREO for the first time give us the possibility to perform remote sensing in white-light between the Sun and the Earth from outside the Sun-Earth line. Because of their wide observation angles (HI1: $4-24^{\circ}$, HI2: $18.7-88.7^{\circ}$) and scattering effects, the interpretation of these images is challenging \citep[\textit{e.g.}][]{vou06,kah07,howtap09}. To derive the kinematics of ICME fronts on their way through the inner heliosphere we used remote observations of both HI instruments on STEREO-A as well as the \textit{in-situ} proton density and velocity delivered by the \textit{Wind} spacecraft near Earth \citep[SWE:][]{ogi95} and STEREO-B \citep[PLASTIC:][]{gal08}.

In this study, we discuss three CME-ICME events observed end-to-end from the Sun to 1~AU, covering a wide range of CME initial conditions and parameters.
The first event of 2\,--\,6 June 2008 is an example of a very slow streamer-blowout type CME with a small longitudinal angular width \citep{moe09apj,rob09,lyn10,woo10}. The CME of 13\,--\,18 February 2009 originated from a bipolar active region and was associated with an EUV wave \citep{kie09,pat09}. It was slow as well but reached its propagation velocity of about 350 km~s$^{-1}$ very close to the Sun \citep[][]{mik11}. The event is expected to have a wider extent in the ecliptic because of a lower axis inclination \citep{moe11}. The third event of 3\,--\,5 April 2010 was the first fast and geoeffective ICME of Solar Cycle 24 \citep{moe10,liu11,woo11} and led to a damage of the Galaxy 15 satellite in geosynchronous orbit and to what has been called a perfect substorm.    

\section{Methods}
\label{s:methods}

For the interpretation of HI images it is necessary to consider scattering effects. ICMEs are detected as the photospheric light scattered off free electrons in the CME body, called Thomson scattering \citep[\textit{e.g.}][]{bil66,hun93}. The scattered light has its maximum when the line of sight (LoS) is perpendicular to the line between the Sun and the scattering particle. That is the reason light that we see in the white-light images originates on a sphere with the Sun-observer line as its diameter, also called the Thomson-surface \citep[TS:][]{vou06}. When leaving the TS the intensity of the scattered light decreases \citep[\textit{e.g.}][]{mor09}. Brightness distribution and morphology are different depending on the ICME axial orientation \citep{cre04}. Thomson scattering influences the geometrical derivations of the following techniques.

Signatures of ICMEs are measured within time-elongation plots \citep[Jmaps:][]{she99,dav09b} which are produced out of stripes along the ecliptic plane extracted from heliospheric images. These stripes are rotated and aligned next to each other with the time increasing in the \textit{x-} and the elongation in the \textit{y-}direction. When an ICME feature is measured within a Jmap the distance from Sun-centre is denoted in elongation, \textit{i.e.}\ in angular degrees. To use the measurements for further analysis we are interested in the ICME kinematics expressed in radial distance, and thus the measured elongation has to be converted. These calculations are limited because of the supposed shape of the ICME front, which cannot be determined by using remote sensing from only one vantage point. These methods and techniques are reviewed and discussed in \citet{liu10b}. In the following calculations we converted the elongation angle into radial distance by using two methods that make different geometrical assumptions for the shape of the ICME front: Fixed-$\phi$ and Harmonic Mean.

\subsection{Fixed-$\phi$ Method}

\begin{figure} 
 \centerline{\includegraphics[width=0.8\textwidth,clip=]{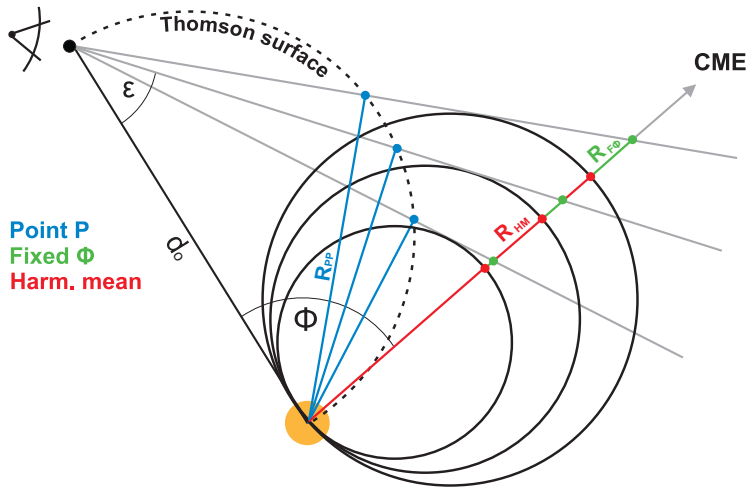}}
 \caption{Comparison between Point P, Fixed-$\phi$, and Harmonic Mean methods. The gray line shows the line of sight, angle $\epsilon$ is the measured elongation, $d_{o}$ is the Sun-observer distance, $\phi$ is the propagation direction relative to $d_{o}$. For the same value of $\phi$, Fixed-$\phi$ gives a larger distance than HM.}
 \label{fig:ppfphihmean}
\end{figure}

The simplest way to convert elongation into distance is the Point P (PP) method \citep{how06}. It assumes a CME as a circle all the way around the Sun. In contrast to PP the Fixed-$\phi$ method \citep[F$\phi$:][]{she99,kah07} assumes a radial propagation of a single plasma element along a straight line (see Figure~\ref{fig:ppfphihmean}), \textit{i.e.}\ a constant propagation direction. This is an important difference to other methods \citep[\textit{e.g.}\ the triangulation technique by][]{liu10a} that make geometrical assumptions of a point wise extent of the CME as well but use two different vantage points (STEREO-A and STEREO-B) and are therefore able to determine the direction for each point along the track. This approach converts the measured elongation angle into radial distance from the Sun for a given propagation direction of the ICME:

\begin{equation}
    R_{F \phi}(t) = d_{o} \frac{\sin \epsilon(t)}{\sin (\epsilon(t) + \phi)},
 \label{eq:fphi}
\end{equation}
where $R_{F\phi}(t)$ is the calculated distance, $d_{o}$ is the Sun-observer distance, $\epsilon(t)$ the measured elongation angle and $\phi$ the derived propagation direction, measured away from the observer, with positive values corresponding to solar West.

\begin{figure} 
 \centerline{\includegraphics[width=\textwidth,clip=]{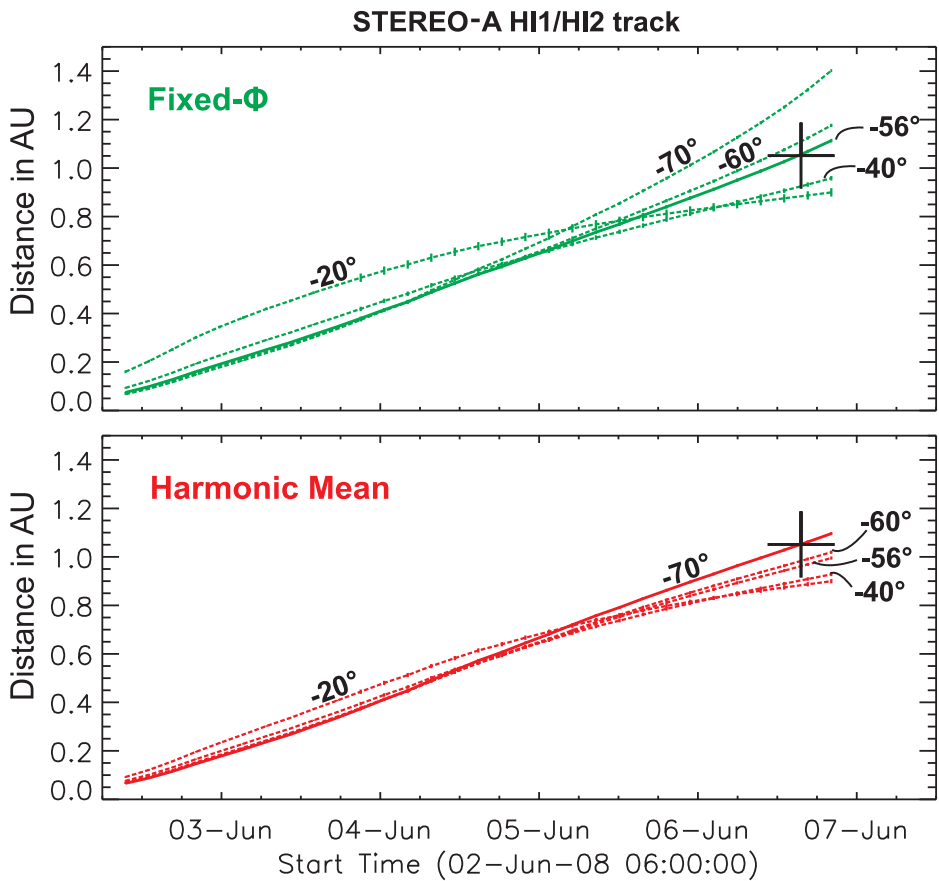}}
 \caption{Resulting radial distance as function of time for five different propagation directions for the F$\phi$ (upper panel) and the HM (lower panel) methods. The angle is the propagation direction relative to the observer --- negative means eastward. The solid horizontal line marks the location of the spacecraft providing the \textit{in-situ} measurements and the vertical solid line indicates the arrival time of the ICME at the \textit{in-situ} spacecraft.}
 \label{fig:angle_calc}
\end{figure}

\subsection{Harmonic Mean Method}

For wide CMEs, the Harmonic Mean method \citep[HM:][]{howtap09,lug09b} may be more appropriate than F$\phi$. It assumes that the measured part is not a single particle but a sphere (or a circle in the ecliptic plane) connected to the Sun at all times. As shown in Figure \ref{fig:ppfphihmean}, it further assumes that the observer always looks along the tangent to this circle. The resulting equation can be understood as the harmonic mean of the PP and F$\phi$ methods:

\begin{equation}
R_{HM}(t) = \frac{2 d_{o} \sin \epsilon(t)}{1 + \sin (\epsilon(t) + \phi)},
\label{eq:hmean1}
\end{equation}
where $R_{HM}(t)$ is the calculated distance, $d_{o}$ is the Sun-observer distance, $\epsilon(t)$ the measured elongation angle, and $\phi$ the derived propagation direction. This was introduced by \cite{lug09b} who found that CME velocities at large elongation angles in a simulation were reproduced better with the HM method, and were over- and underestimated by the Fixed-$\Phi$ and Point~P methods, respectively.

\subsection{Combining Remote and \textit{In-Situ} Signals}

We have developed a new technique, based on the two conversion methods, by adding two boundary conditions:
first, the arrival time [$t_{a}$] of the ICME front at the \textit{in-situ} spacecraft and second, the \textit{in-situ} measured velocity [$V_{i}$]. As an illustration, Figure \ref{fig:angle_calc} shows the resulting distance-time profiles of the 1\,--\,6 June 2008 event calculated with different propagation angles (Top: F$\phi$, Bottom: HM). The cross is the measured arrival time of the ICME front at the location of the \textit{in-situ} spacecraft.

\subsubsection{Boundary Condition: \textit{In-Situ} Arrival Time}

For estimating the most reliable propagation angle [$\phi$] the measured elongation value at the \textit{in-situ} arrival time of the ICME [$\epsilon_{t_{a}}$] is converted with F$\phi$ into a distance, $R_{F\phi,t_{a}}(\phi)$, by using different angles ($\phi \in [0^{\circ},180^{\circ} ]$). $R_{F\phi,t_{a}}(\phi)$ is subtracted from the distance between the Sun and the \textit{in-situ} spacecraft [$d_{i}$]:

\begin{equation}
\Delta d_{F\phi}(\phi) = R_{F\phi,t_{a}}(\phi) - d_{i}.
\label{eq:rfphi}
\end{equation}
The angle which minimizes $\Delta d_{F\phi}(\phi)$, which we call $\phi_{dF\phi}$, is the resulting direction from the constraint with the arrival time. 

\begin{figure} 
 \centerline{\includegraphics[width=0.8\textwidth,clip=]{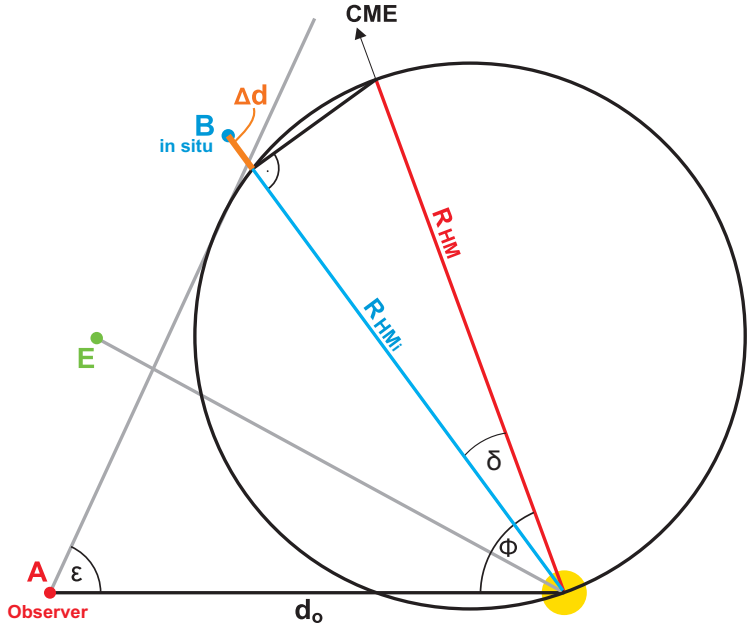}}
 \caption{Illustration showing the calculation of the distance [$\Delta d$] from the Sun to the point of the circular CME front in the direction of the \textit{in-situ} observing spacecraft. The gray line shows the line of sight, angle $\epsilon$ is the measured elongation, $d_{o}$ is the Sun-observer distance, $\phi$ is the propagation direction relative to $d_{o}$, $R_{HM}$ is the distance of the CME apex from the Sun, $R_{HMi}$ is the distance of the CME front in direction of the \textit{in-situ} spacecraft from the Sun, and $\delta$ is the angle between $R_{HM}$ and $R_{HMi}$.}
 \label{fig:hm_insitu}
\end{figure}

To do the same calculation for HM we have to consider the particular, circular geometry of the front. To this end one cannot use the whole diameter of the circle [$R_{HM}$], but rather the distance between the Sun and the point of intersection of the HM circle with the line connecting the Sun and the \textit{in-situ} spacecraft, as illustrated in Figure~\ref{fig:hm_insitu}. We call this distance $R_{HMi,t_{a}}(\phi)$, and it is calculated by 

\begin{equation}
R_{HMi,t_{a}}(\phi) = R_{HM,t_{a}}(\phi)\cos{\delta},
\label{eq:cosinus}
\end{equation}
where $\delta$ is the angle between $d_{i}$ and $R_{HM}$. Similar to above, we calculate:

\begin{equation}
\Delta d_{HMi}(\phi) = R_{HMi,t_{a}}(\phi) - d_{i},
\label{eq:rinsitu}
\end{equation}
where $\Delta d_{HMi}(\phi)$ is the resulting difference and $R_{HMi,t_{a}}(\phi)$ the result of HM along $d_{i}$. The propagation direction [$\phi_{dHM}$] corresponding to the minimum of $\Delta d_{HMi}(\phi)$, is again the outcome of the technique.

\subsubsection{Boundary Condition: Velocity at \textit{In-Situ} Arrival Time}

We applied the same minimization method as described before by subtracting the \textit{in-situ} measured velocity [$V_{i}$] from the calculated velocity converted from the measured elongation value at $t_{a}$, $V_{F\phi,t_{a}} (\phi \in [0^{\circ},180^{\circ} ]$):

\begin{equation}
\Delta V_{F\phi}(\phi) = V_{F\phi,t_{a}}(\phi) - V_{i},
\label{eq:vfphiinsitu}
\end{equation}
where $\Delta V_{F\phi}(\phi)$ is the calculated difference, $V_{F\phi,t_{a}}(\phi)$ the derived velocity at arrival time, and $V_{i}$ is the \textit{in-situ} measured velocity of the ICME front.

\begin{figure} 
 \centerline{\includegraphics[width=0.8\textwidth]{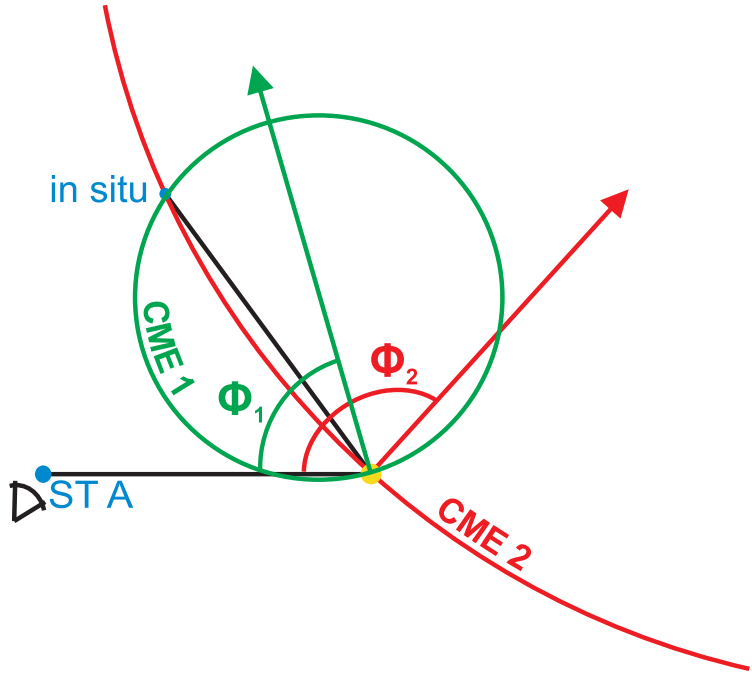}}
 \caption{Two possible solutions for HM. The green arrow shows the direction of the ICME apex resulting from the first minimum and the green circle shows the ICME front (CME~1) for this direction at arrival time. The red arrow indicates the direction derived from the second minimum of $\Delta V$ for the ICME event of June 2008. The apex of this ICME (CME~2) would be $\approx 2.5$ AU away from the Sun at arrival time at the \textit{in-situ} spacecraft.}
 \label{fig:minima}
\end{figure}

The same method can be applied by using HM. The velocity $V_{HMi}$ was calculated on the basis of $R_{HMi}$, thus in the direction of the \textit{in-situ} spacecraft, which for HM does not necessarily have to be the apex of the ICME front, but can, in principle, be also any point along the circle. We define

\begin{equation}
\Delta V_{HMi}(\phi) = V_{HMi,t_{a}}(\phi) - V_{i},
\label{eq:vhminsitu}
\end{equation}
where $\Delta V_{HMi}(\phi)$ is the calculated difference and $V_{HMi,t_{a}}(\phi)$ the derived velocity at arrival time. 
The angle belonging to the minimum difference [$\phi_{VF\Phi}$ or $\phi_{VHM}$] is the direction resulting from the constraint with the velocity at arrival time. \citet{liu10b} pointed out that there is an ambiguity for possible propagation directions for HM. Because of mathematical reasons there are two minima within the range of $\phi \in [0^\circ,180^\circ]$. Since we get two minima for both constraints ($\Delta d$, $\Delta V$) we choose the value where both minima show the best agreement with each other; in our cases, it seems the smaller value. The conversion using the next minimum would yield a large distance of the CME apex, which may not be possible given the observed CME size from two vantage points (Figure~\ref{fig:minima}).

The directions from the arrival time and velocity constraints will not be identical. To combine the resulting propagation directions for each method, we define

\begin{eqnarray}
 \phi_{F\phi} &=& (\phi_{dF\phi}+ \phi_{VF\phi})/2\\
 \phi_{HM}    &=& (\phi_{dHM}+ \phi_{VHM})/2  
\end{eqnarray}
so we take the average values for the direction for each method and use these for further analysis. Here, $\phi_{HM}$ is the propagation direction of the ICME apex for the HM method. However, all kinematics are calculated for the part of the circle which hit the \textit{in-situ} spacecraft and not for the apex.

The distance-time profile for the defined mean propagation angle was fitted by using a cubic spline, from which we derived the velocity profile via numerical differentiation. All errors result from the manual tracking of the ICME front only. To be able to do an error estimation, every feature was measured five times. The resulting standard deviation also yields a reliable error for the numerical velocity derivation  using three-point, Lagrangian interpolation. We also indicate an error for the propagation direction for both methods.

\section{Results}
\label{s:results}

Three different ICMEs were investigated to cover a range from slow to fast CME initial speeds. One event was slow and narrow (June 2008), the second was also slow but had a wide longitudinal extent (February 2009), and the third ICME (April 2010) was a fast and geoeffective event.

\subsection{02\,--\,06 June 2008 ICME}

\begin{figure}
 \centerline{\includegraphics[width=\textwidth,clip=]{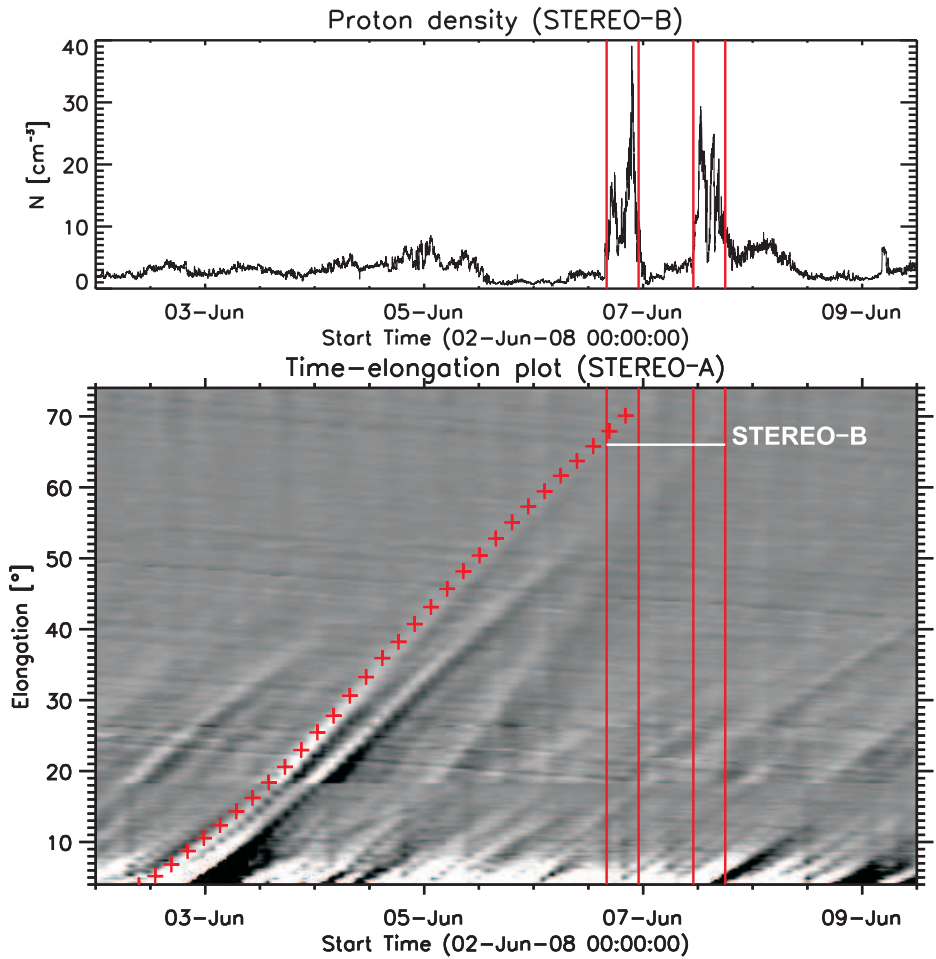}}
 \caption{\textit{In-situ} measured proton density from STEREO-B (top). The first red line from the left marks the time of the ICME arriving at STEREO-B and others delimit two strong peaks in the proton density. The lower panel shows the Jmap produced from remote sensing data of heliospheric images of STEREO-A with overplotted measurement points (red crosses). The white horizontal line indicates the position of STEREO-B.}
 \label{fig:jun08_is_jplot}
\end{figure}

The leading edge of this ICME was seen on 2 June 2008 09:00 UT in the HI1 field of view (FoV) and on 3 June 2008 12:00 UT in the HI2 FoV of STEREO-A. This CME had no obvious signatures of magnetic reconnection on the Sun \citep{rob09}. A magnetic cloud was detected between 6 June 2008 $\approx$~22~UT and 7 June 2008 $\approx$~12~UT at STEREO-B \citep{moe09apj,lyn10,woo10}. The separation between both STEREO spacecraft was $53.9^\circ$ at that time.
Figure \ref{fig:jun08_is_jplot} shows the proton density (STEREO-B, PLASTIC) and the Jmap (STEREO-A, HI1/2) with two clear traces of the event. The first one is the leading edge and the second one is the core of the ICME. Both Jmap-tracks match with the \textit{in-situ} arrival times (sudden increase of the proton density at STEREO-B) as indicated with the vertical red lines. Its \textit{in-situ} signature as well as the heliospheric images both reveal a pronounced three-part structure, with the density enhancements bracketing the magnetic flux rope \citep{moe09apj,lyn10}.

\begin{figure}
 \centerline{\includegraphics[width=\textwidth,clip=]{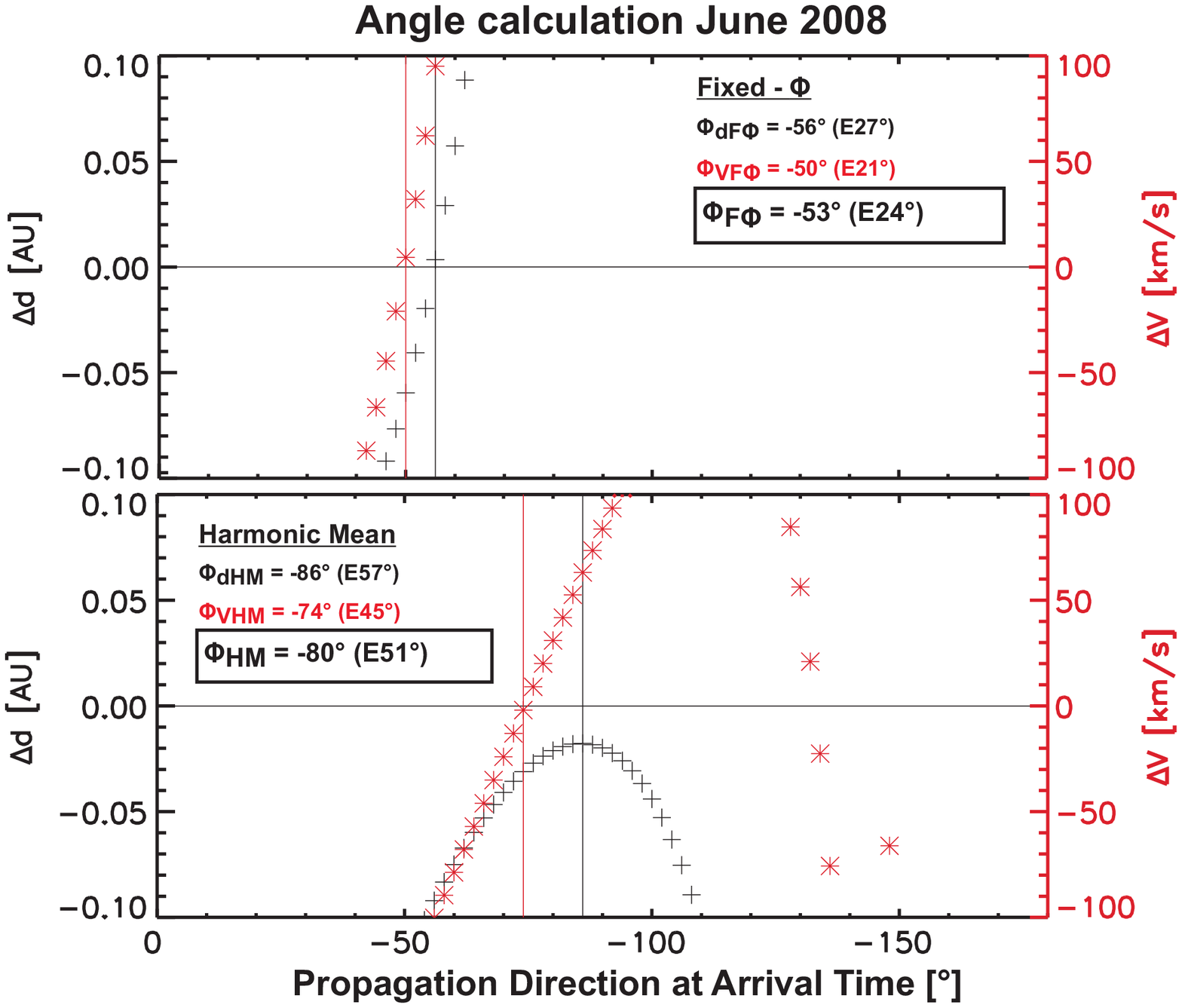}}
 \caption{Propagation directions for Fixed-$\phi$ (top panel) and Harmonic Mean (bottom panel). The black crosses show the differences between the calculated distance at arrival time and the distance of the \textit{in-situ} spacecraft from Sun-centre for different propagation angles. The red asterisks indicate the same approach but for the difference of the calculated velocity and the \textit{in-situ} measured velocity at arrival time.}
 \label{fig:jun08_angle_calc}
\end{figure}

To obtain the propagation direction and the velocity only the first track, \textit{i.e.}\ the ICME leading edge, was investigated. Figure \ref{fig:jun08_angle_calc} shows the calculation of the minima of the differences between the converted distance at \textit{in-situ} arrival time and the distance between Sun and \textit{in-situ} spacecraft ($\phi_{dF\phi}$ and $\phi_{dHM}$) as well as for the differences between the derived velocity at arrival time and the \textit{in-situ} measured arrival velocity [$\phi_{VF\phi}$ and $\phi_{VHM}$]. F$\phi$ (top panel) yields as a mean direction E$24 \pm 3^\circ$ (relative to the Sun-Earth line) and HM (bottom panel) results in E$51 \pm 6^\circ$.

\begin{figure} 
 \centerline{\includegraphics[width=\textwidth,clip=]{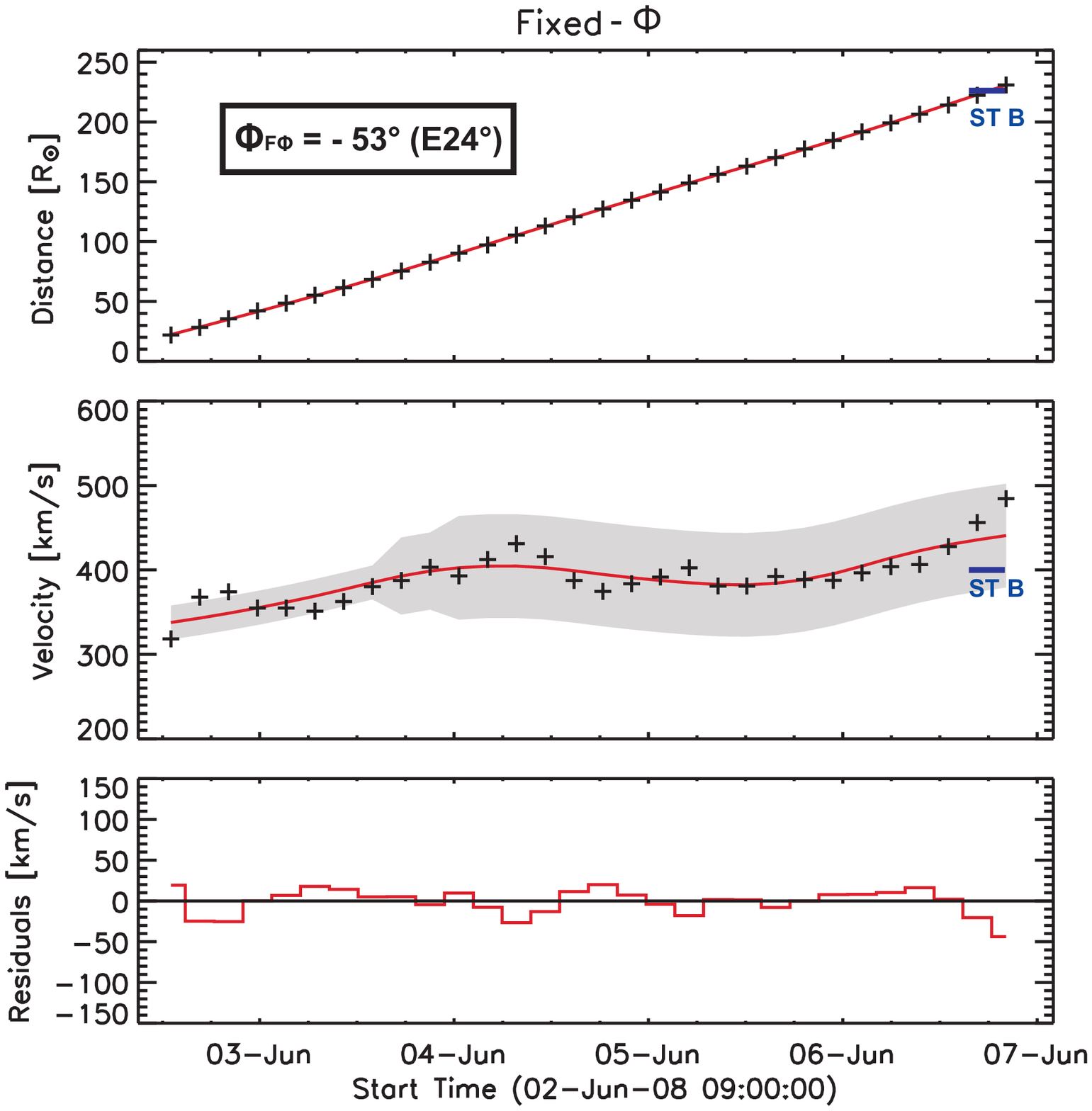}}
 \caption{Top: Resulting distance-time profile for F$\phi$. The crosses are the converted mean values of the direct measurements and the red line is the spline fit. The blue horizontal line shows the arrival time of the ICME at STEREO-B. Middle: The solid curve shows the derivation of the fit, which is done to determine the velocities of this event. The gray area indicates the standard deviation of the measurements. The horizontal line indicates the \textit{in-situ} measured velocity at STEREO-B. Bottom: Residuals of the fit and the direct measurements.}
 \label{fig:jun08_FPH}
\end{figure}

\begin{figure} 
 \centerline{\includegraphics[width=\textwidth,clip=]{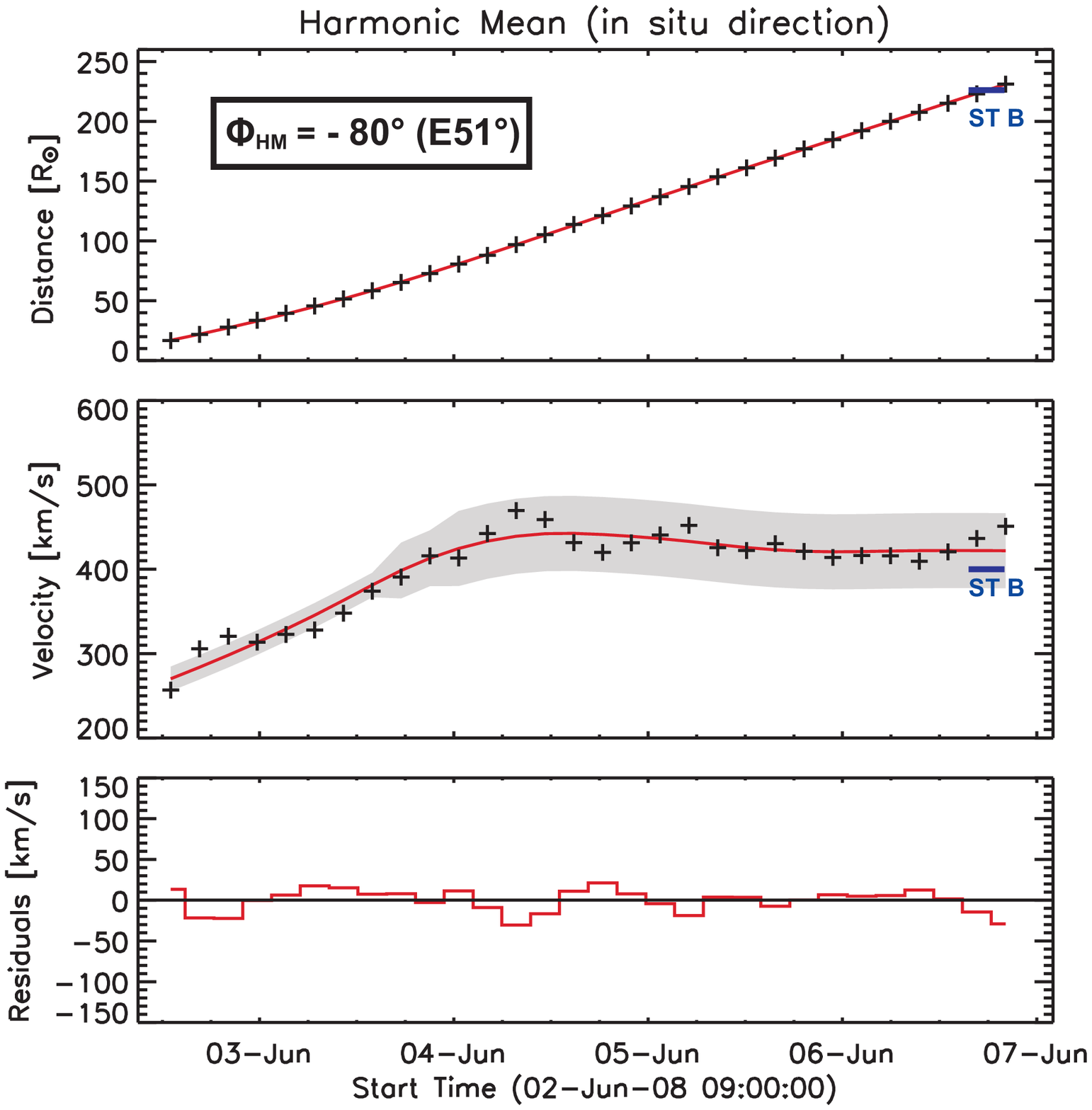}}
 \caption{Top: Resulting distance-time profile for HM. The crosses are the converted mean values of the direct measurements and the red line is the spline fit. The blue horizontal line shows the arrival time of the ICME at STEREO-B. Middle: The solid curve shows the derivation of the fit, which is done to determine the velocities of this event. The gray area indicates the standard deviation of the measurements. The horizontal line indicates the \textit{in-situ} measured velocity at STEREO-B. Bottom: Residuals of the fit and the direct measurements.}
 \label{fig:jun08_HM}
\end{figure}

Figure \ref{fig:jun08_FPH} shows the kinematics of this event derived from F$\phi$ using the obtained propagation angle. The upper panel displays the result of the conversion from elongation into solar radii and the link to the arrival time at the \textit{in-situ} spacecraft. The standard deviation lies between 0.5 R$_{\odot}$ (HI1) and 1.6 R$_{\odot}$ (HI2). The middle panel shows the direct derivation of the measurements (crosses) and the derivative of the spline fit. The kinematics using F$\phi$ ($\phi_{F\phi}=$  E$24 \pm 3^\circ$) yield a continuous acceleration from $\approx 330$ to $\approx 440$ km~s\textsuperscript{$-1$} and a mean velocity of $389 \pm 48$~km~s\textsuperscript{$-1$}. The velocity at arrival at STEREO-B derived from F$\phi$ is $432 \pm 61$~km~s\textsuperscript{$-1$}. F$\phi$ results in a typical late acceleration at $\approx$ 1~AU. This can not be interpreted as a true behavior of the CME but is an artefact of the method \citep[cf. the discussions in][and]{lug09b} and \citet{woo09}. The new method using \textit{in-situ} data as constraints diminishes this effect. The same approach but for HM is shown in Figure \ref{fig:jun08_HM}. The errors of the measurements are between 0.4 R$_{\odot}$ (HI1) and 1.2 R$_{\odot}$ (HI2). Kinematics using HM ($\phi_{HM}= $  E$51 \pm 6^\circ$) deliver a mean velocity of $395 \pm 35$~km~s\textsuperscript{$-1$}. The CME shows a strong acceleration from $\approx 280$ to $\approx 440$~km~s\textsuperscript{$-1$} up to a distance of about 100 solar radii followed by a slight deceleration to a final velocity of $422 \pm 44$~km~s\textsuperscript{$-1$}, consistent with the \textit{in-situ} measured velocity at STEREO-B of 403~km~s\textsuperscript{$-1$}. 

\subsection{13\,--\,18 February 2009 ICME}

The ICME arrived on 13 February 2009 11:00 UT in the HI1 FoV of STEREO-A and became visible in HI2 on 14 February 2009 04:00 UT. This CME was associated with a flare and an EUV wave \citep{kie09} occurring at the limb as seen by STEREO-A, since the separation of the STEREO satellites was $91^\circ$ at that time. Compared to the June 2008 event there is no characteristic three-part structure --- neither in the heliospheric images nor in the \textit{in-situ} data. STEREO-B measured an increase of the proton density that was followed by a large-scale magnetic flux rope \citep{moe11}.
\begin{figure} 
 \centerline{\includegraphics[width=\textwidth,clip=]{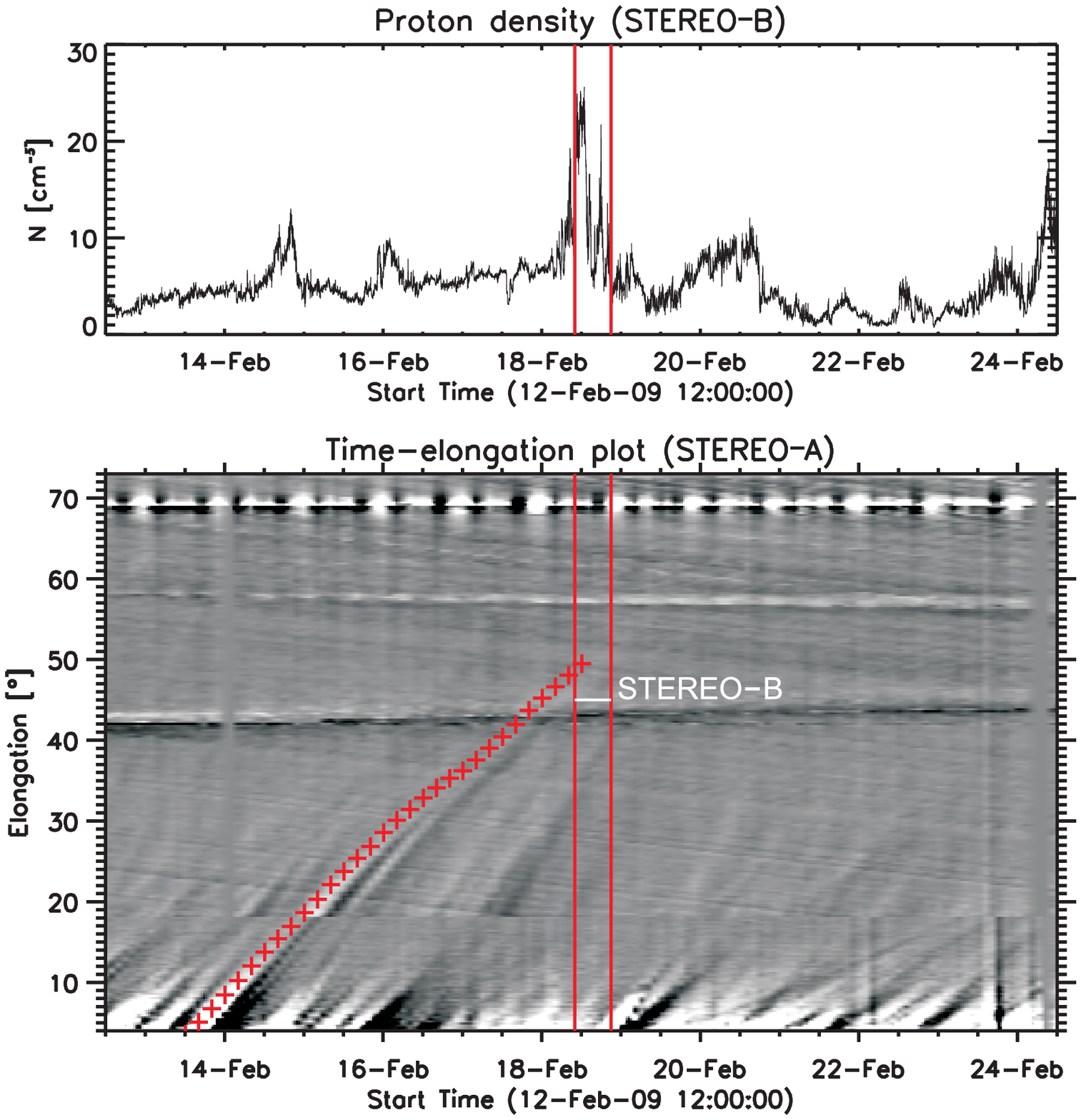}}
 \caption{\textit{In-situ} measured proton density from STEREO-B (top). The first red line from the left marks the time of the ICME arriving at STEREO-B and others delimit two strong peaks in the proton density. The lower panel shows the Jmap produced from remote sensing data of heliospheric images of STEREO-A with overplotted measurement points (red crosses). The white horizontal line indicates the position of STEREO-B.}
 \label{fig:feb09_is_jplot}
\end{figure}

\begin{figure}
 \centerline{\includegraphics[width=\textwidth,clip=]{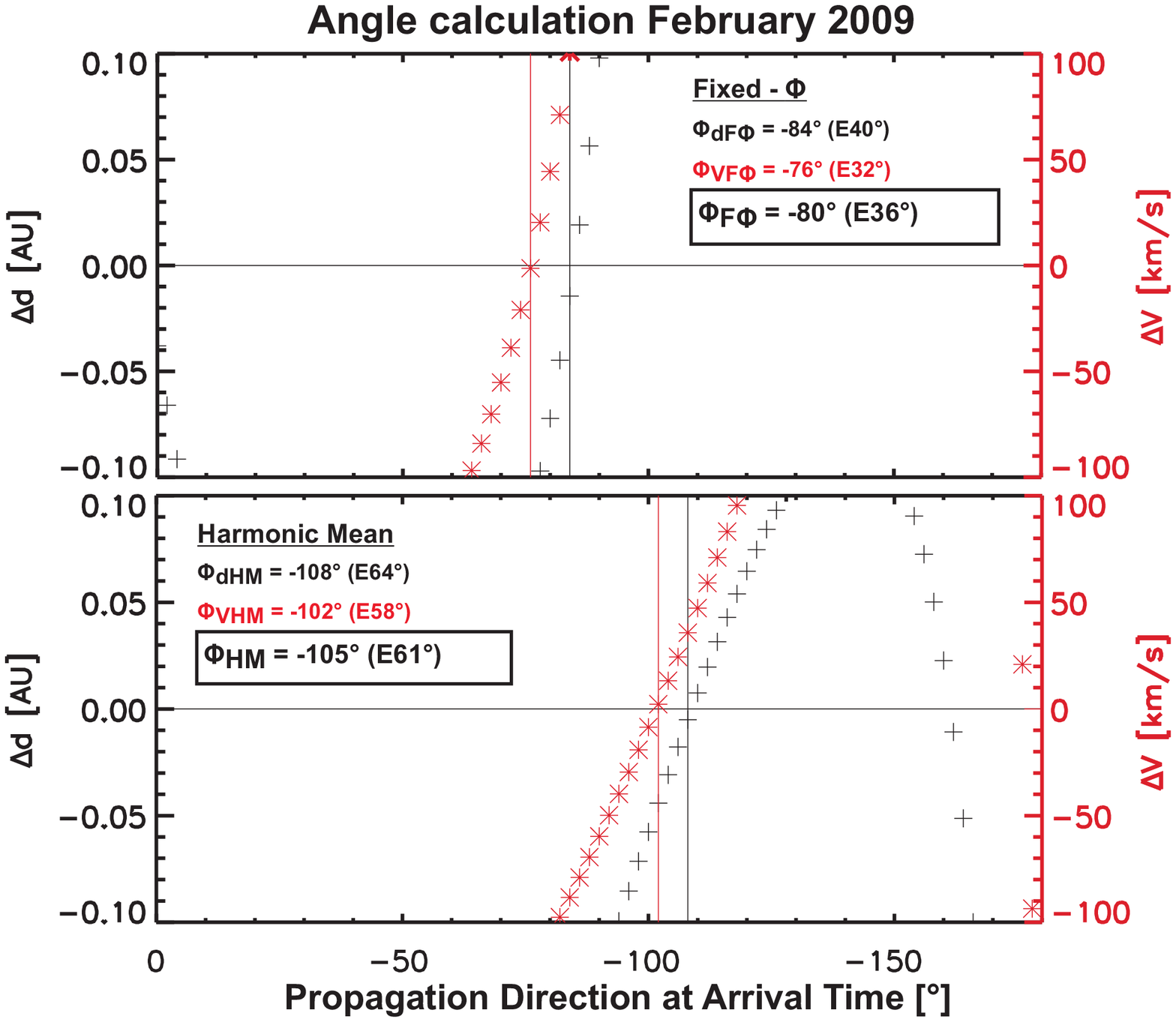}}
 \caption{Propagation directions for Fixed-$\phi$ (top panel) and Harmonic Mean (bottom panel). The black crosses show the differences between the calculated distance at arrival time and the distance of the \textit{in-situ} spacecraft from Sun-centre for different propagation angles. The red asterisks indicate the same approach but for the difference of the calculated velocity and the \textit{in-situ} measured velocity at arrival time.}
 \label{fig:feb09_angle_calc}
\end{figure}

Figure \ref{fig:feb09_is_jplot} shows the proton density measured by STEREO-B and the Jmap from STEREO-A. The ICME track became faint rather quickly, as expected for a limb CME \citep{mor09}. Comparing the track in the Jmap and the proton density in Figure \ref{fig:feb09_is_jplot} shows that the ICME seems to arrive earlier at the elongation of STEREO-B than it was measured \textit{in-situ}.

\begin{figure} 
 \centerline{\includegraphics[width=\textwidth,clip=]{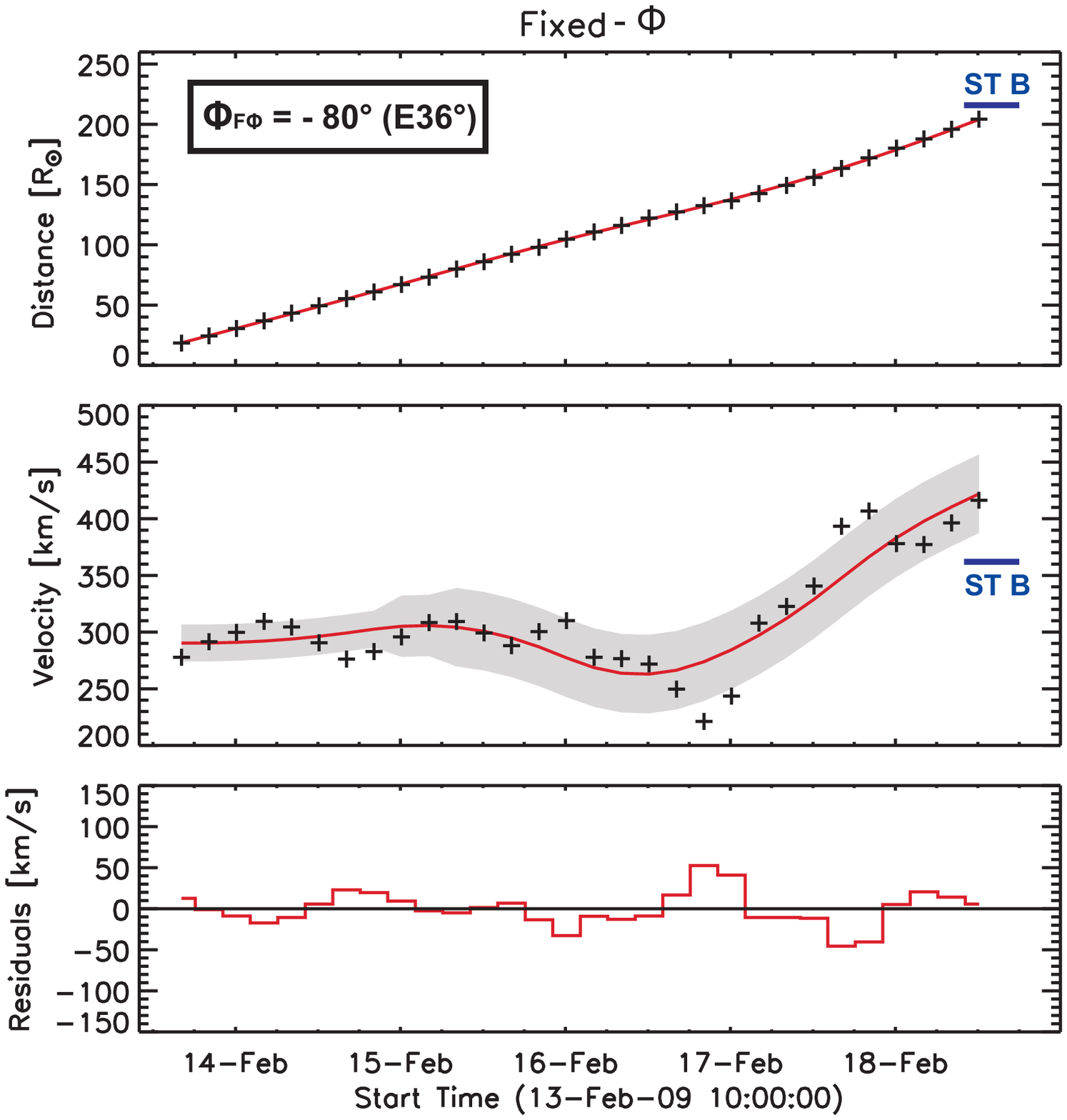}}
 \caption{Top: Resulting distance-time profile for F$\phi$. The crosses are the converted mean values of the direct measurements and the red line is the spline fit. The blue horizontal line shows the arrival time of the ICME at STEREO-B. Middle: The solid curve shows the derivation of the fit, which is done to determine the velocities of this event. The gray area indicates the standard deviation of the measurements. The horizontal line indicates the \textit{in-situ} measured velocity at STEREO-B. Bottom: Residuals of the fit and the direct measurements.}
 \label{fig:feb09_FPH}
\end{figure}

\begin{figure} 
 \centerline{\includegraphics[width=\textwidth,clip=]{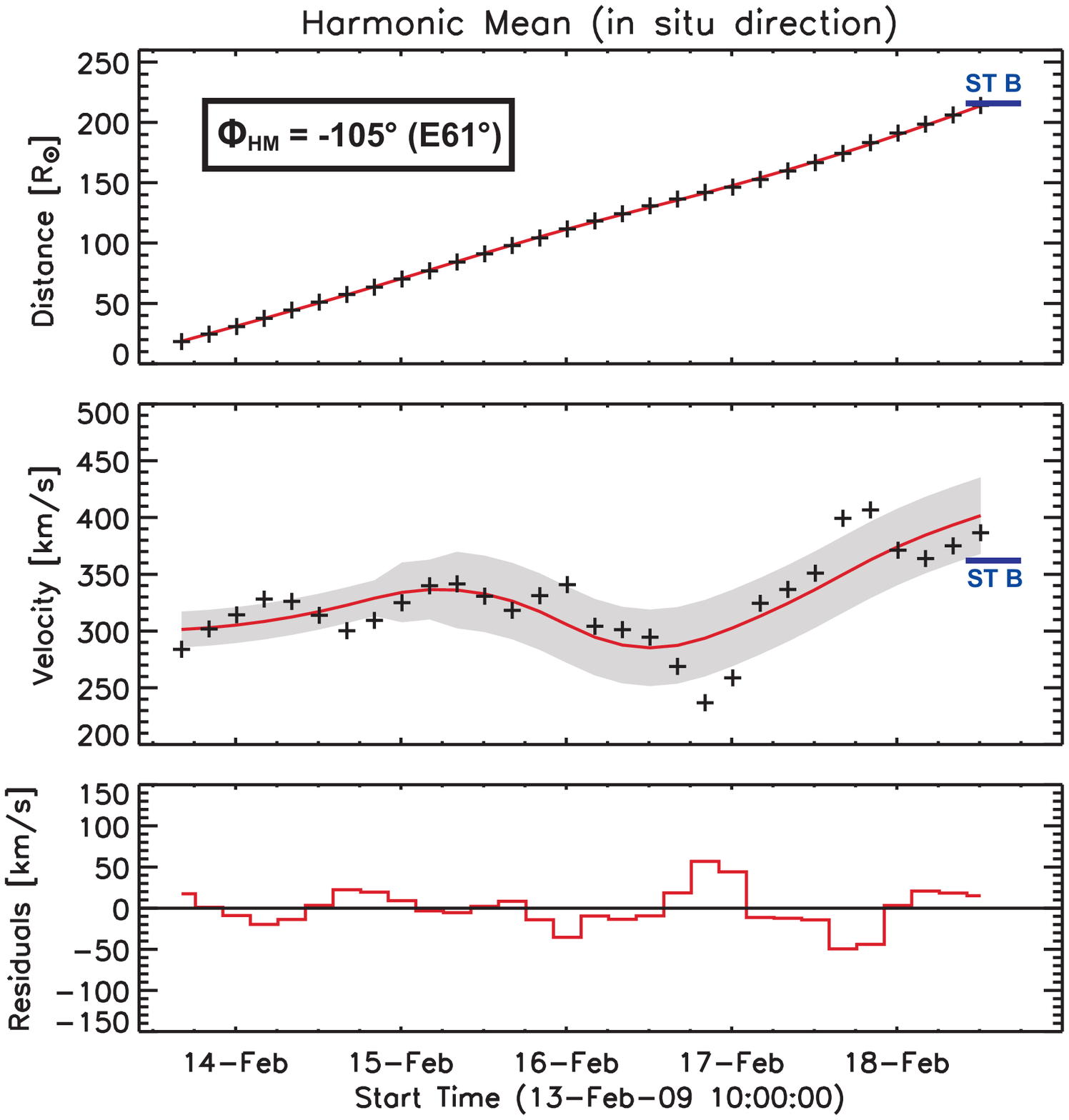}}
 \caption{Top: Resulting distance-time profile for HM. The crosses are the converted mean values of the direct measurements and the red line is the spline fit. The blue horizontal line shows the arrival time of the ICME at STEREO-B. Middle: The solid curve shows the derivation of the fit, which is done to determine the velocities of this event. The gray area indicates the standard deviation of the measurements. The horizontal line indicates the \textit{in-situ} measured velocity at STEREO-B. Bottom: Residuals of the fit and the direct measurements.}
 \label{fig:feb09_HM}
\end{figure}

Figure \ref{fig:feb09_angle_calc} illustrates the calculation of the propagation directions ($\phi_{F\phi}=$  E$36 \pm 4^\circ$, $\phi_{HM}=$
 E$61 \pm 3^\circ$). 
Again, HM gives a direction further away from the observer. The upper panel in Figure \ref{fig:feb09_FPH} shows the result of the conversion from elongation into radial distance [F$\phi$] and the linkage with the \textit{in-situ} arrival time. The standard deviation of both conversion methods is between 0.5 R$_{\odot}$ (HI1) and 1 R$_{\odot}$ (HI2). Kinematics using F$\phi$ yield a mean velocity of $310 \pm 29$ km~s\textsuperscript{$-1$}. The velocity profile shows a nearly constant speed up to a distance of $\approx$ 100 R$_\odot$ followed by an acceleration from $\approx 280$ km~s\textsuperscript{$-1$} to a velocity at arrival time of $416 \pm 35$ km~s\textsuperscript{$-1$}. Again F$\phi$ shows an apparent acceleration close to 1~AU. The mean velocity derived of HM is $325 \pm 28$ km~s\textsuperscript{$-1$} and the impact velocity is $398 \pm 34$ km~s\textsuperscript{$-1$}, whereas the \textit{in-situ} measured impact velocity of the ICME at STEREO-B is $\approx 362$ km~s\textsuperscript{$-1$} (see Figure \ref{fig:feb09_HM}). There seems to be an upward kink in the track in the Jmap at about $35^\circ$ elongation that is also visible in the derived kinematics as an acceleration.
%This event was also investigated by \citet{tem11} who compared the ambient solar wind characteristics as derived from ENLIL MHD modelling to the ICME velocity profile measured in the HI images. They found an apparent interaction with a high-speed solar wind stream, which may provide a reliable explanation for the acceleration at $\approx$ 150 R$_{\odot}$.

\subsection{03\,--\,05 April 2010 ICME}

Compared to the previous mentioned events this CME was relatively fast for solar minimum. The ICME arrives in the HI1 FoV of STEREO-A on 3 April 2010 10:00 UT and a shock was detected on 5 April 2010 07:58 UT at \textit{Wind}, situated at the L1 point sunward of the Earth \citep{moe10}. The separation between STEREO-A and \textit{Wind} was 67.4$^\circ$ at that time. It was a geoeffective ICME and caused a moderate geomagnetic storm with a minimum of the Dst index about $-72$ nT and maximum Kp of 8 \citep{moe10,woo11}.

\begin{figure} 
 \centerline{\includegraphics[width=\textwidth,clip=]{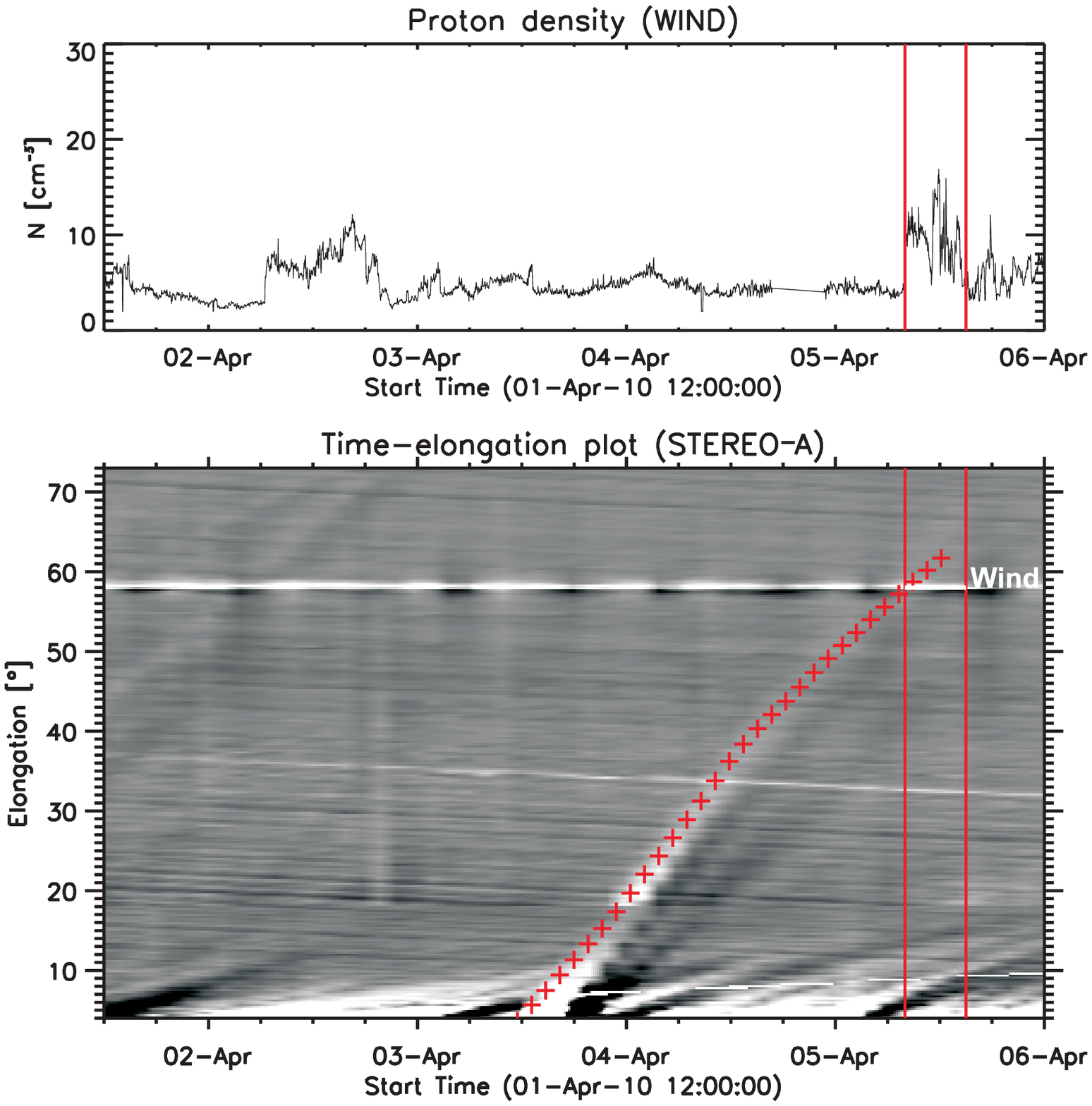}}
 \caption{\textit{In-situ} measured proton density from \textit{Wind} (top). The first red line from the left marks the time of the ICME arriving at \textit{Wind} and others delimit two strong peaks in the proton density. The lower panel shows the Jmap produced from remote sensing data of heliospheric images of STEREO-A with overplotted measurement points (red crosses). The white horizontal line indicates the position of \textit{Wind}.}
 \label{fig:apr10_is_jplot}
\end{figure}

\begin{figure}
 \centerline{\includegraphics[width=\textwidth,clip=]{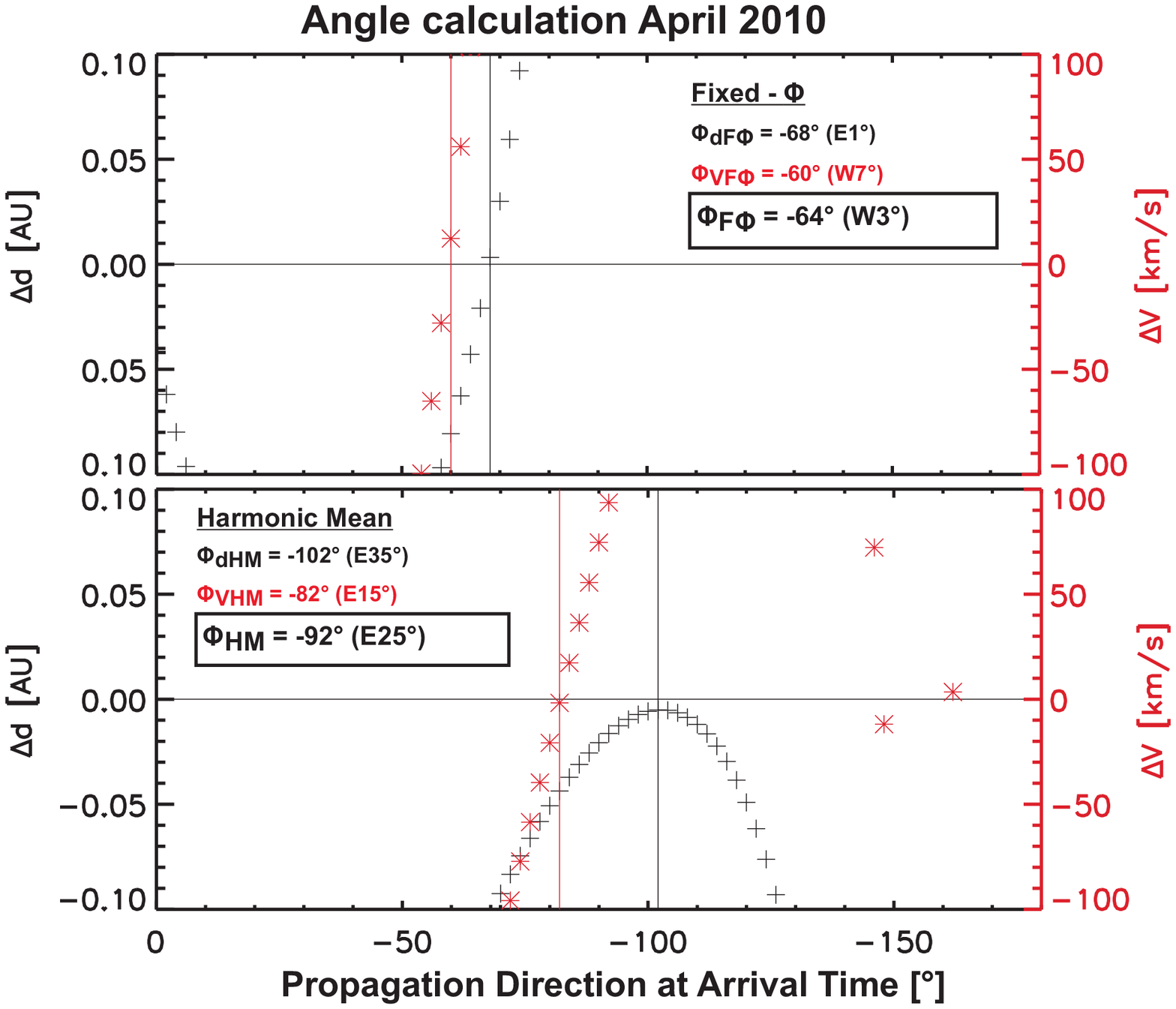}}
 \caption{Propagation directions for Fixed-$\phi$ (top panel) and Harmonic Mean (bottom panel). The black crosses show the differences between the calculated distance at arrival time and the distance of the \textit{in-situ} spacecraft from Sun-centre for different propagation angles. The red asterisks indicate the same approach but for the difference of the calculated velocity and the \textit{in-situ} measured velocity at arrival time.}
 \label{fig:apr10_angle_calc}
\end{figure}

\begin{figure} 
 \centerline{\includegraphics[width=\textwidth,clip=]{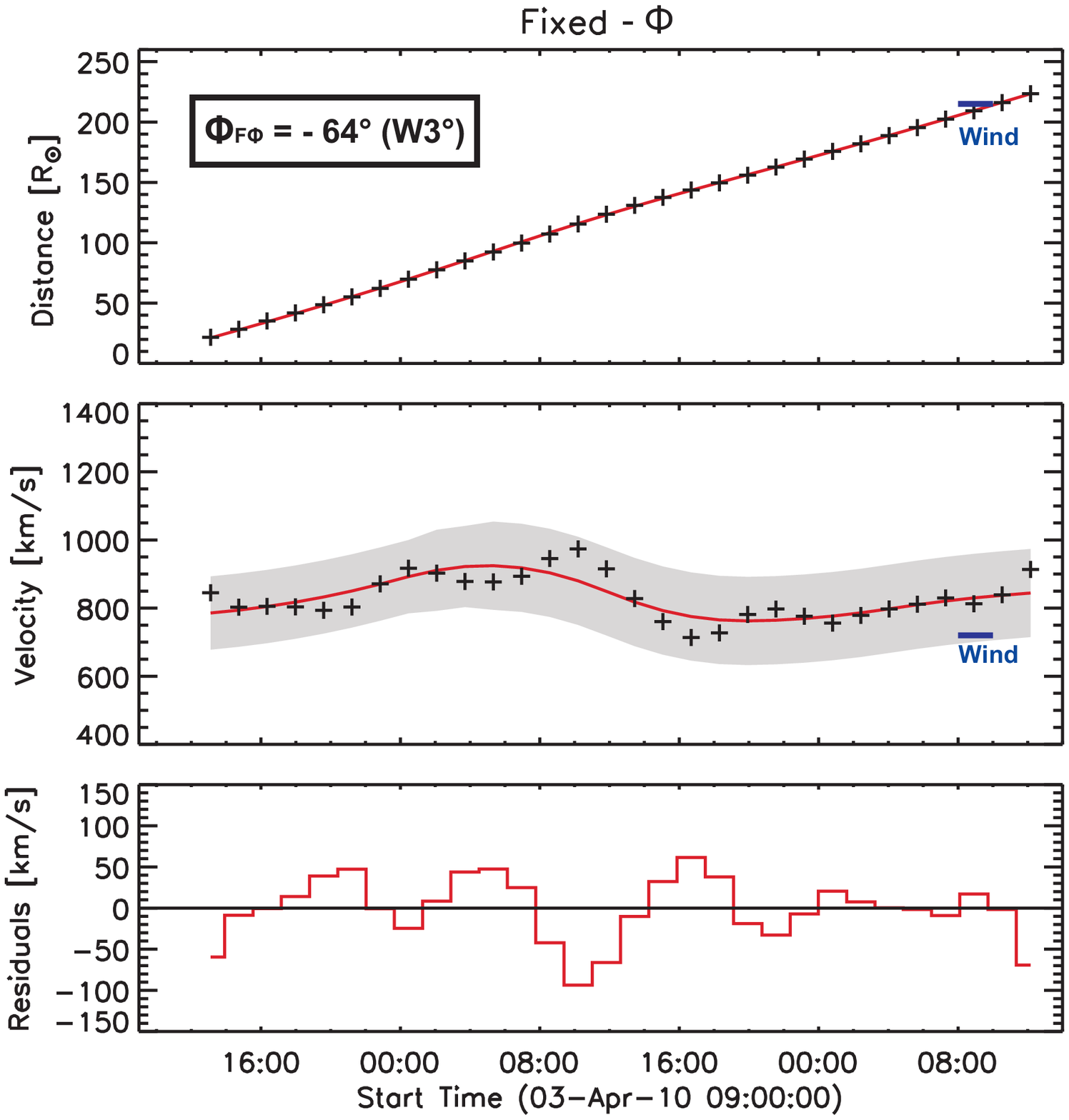}}
 \caption{Top: Resulting distance-time profile for F$\phi$. The crosses are the converted mean values of the direct measurements and the red line is the spline fit. The blue horizontal line shows the arrival time of the ICME at \textit{Wind}. Middle: The solid curve shows the derivation of the fit, which is done to determine the velocities of this event. The gray area indicates the standard deviation of the measurements. The horizontal line indicates the \textit{in-situ} measured velocity at \textit{Wind}. Bottom: Residuals of the fit and the direct measurements.}
 \label{fig:apr10_FPH}
\end{figure}

\begin{figure} 
 \centerline{\includegraphics[width=\textwidth,clip=]{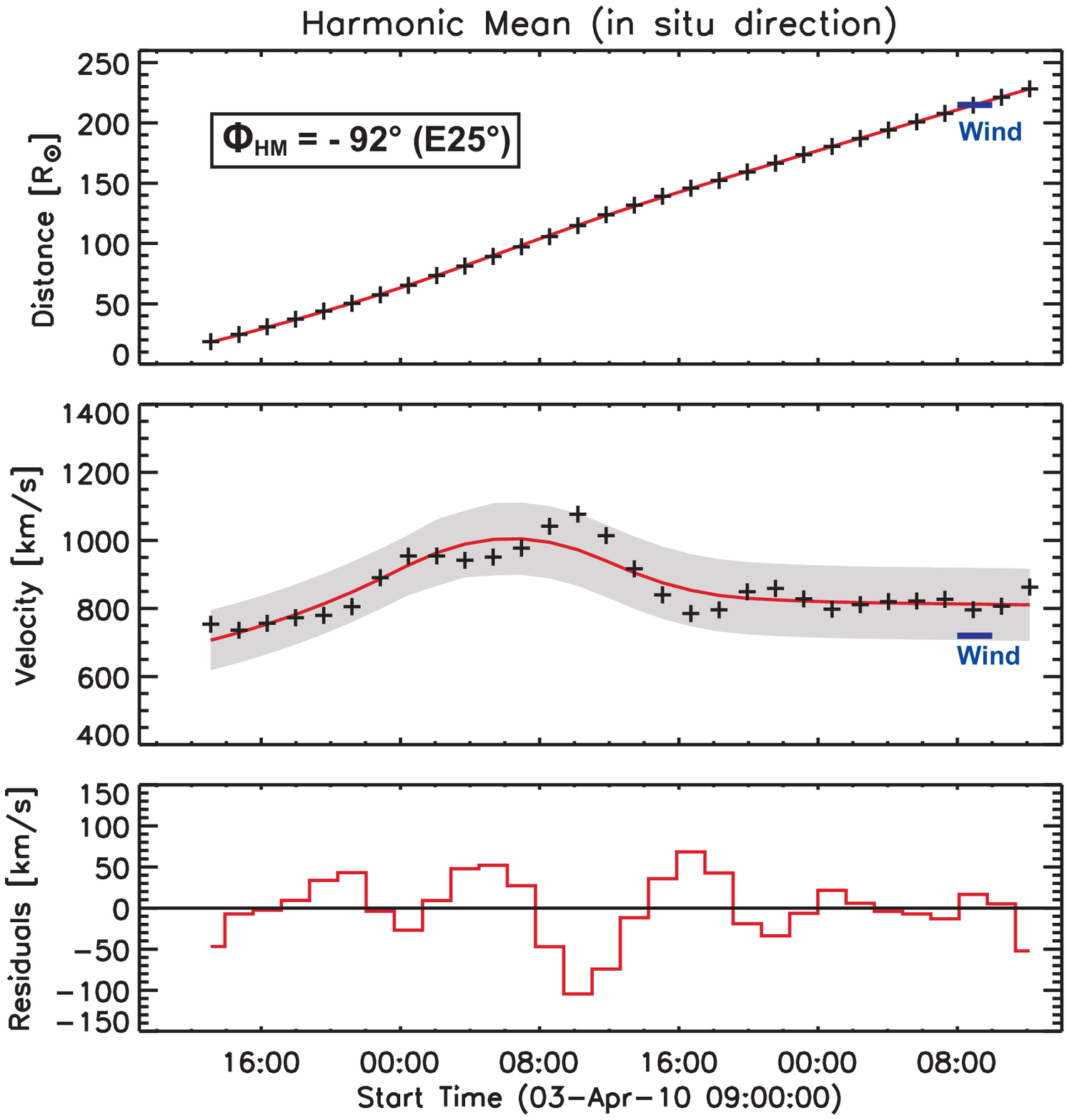}}
 \caption{Top: Resulting distance-time profile for HM. The crosses are the converted mean values of the direct measurements and the red line is the spline fit. The blue horizontal line shows the arrival time of the ICME at \textit{Wind}. Middle: The solid curve shows the derivation of the fit, which is done to determine the velocities of this event. The gray area indicates the standard deviation of the measurements. The horizontal line indicates the \textit{in-situ} measured velocity at \textit{Wind}. Bottom: Residuals of the fit and the direct measurements.}
 \label{fig:apr10_HM}
\end{figure}

Figure \ref{fig:apr10_is_jplot} shows the proton density at \textit{Wind} and the Jmap of HI1A and HI2A of the event. The track of the ICME fits well with the sharp increase in proton density. As illustrated in Figure \ref{fig:apr10_angle_calc} the two conversion methods yield a propagation direction of $\phi_{F\phi}=$ W$3 \pm 4^\circ$ and $\phi_{HM}=$ E$25 \pm 10^\circ$, respectively. The measurement errors for F$\phi$ are 1.3 R$_\odot$ (HI1) and 1.5 R$_\odot$ (HI2), for HM 1 R$_\odot$ (HI1) and 1.3 R$_\odot$ (HI2). Kinematics from F$\phi$ result in a mean velocity of $829 \pm 122$ km~s\textsuperscript{$-1$} and an impact velocity of $825 \pm 129$ km~s\textsuperscript{$-1$} (Figure \ref{fig:apr10_FPH}), kinematics from HM yield a mean velocity of $854 \pm 100$ km~s\textsuperscript{$-1$} and at arrival at \textit{Wind} a velocity of $813 \pm 106$ km~s\textsuperscript{$-1$} (Figure \ref{fig:apr10_HM}). The average speed of the ICME sheath region between the shock and the magnetic cloud as measured \textit{in-situ} by \textit{Wind}, was $\approx$ 720 km~s\textsuperscript{$-1$}. Both methods slightly overestimate the impact velocity by about 100 km~s\textsuperscript{$-1$}. In this case, differences in the velocity-time profiles are more pronounced: F$\phi$ shows a nearly constant velocity between 800 and 900 km~s\textsuperscript{$-1$} while HM delivers an acceleration up to $\approx 100$ solar radii to about 1000 km~s\textsuperscript{$-1$} followed by a deceleration to 800 km~s\textsuperscript{$-1$}. Since the irregularity in the velocity profile happens in the transition region between the fields of view of HI1 and HI2 it could also be related to the different sensitivities of the cameras. The signal of HI1 turns faint rather quickly at larger elongations while HI2 seems to overexpose at lower elongations what yields to a discontinuity of the front and makes the tracking of the leading edge difficult. These circumstances could be a possible reason for the untypical shape of the velocity evolution.

Table \ref{tab:results} lists the results of the three events for the two conversion methods, F$\phi$ and HM, and the relevant \textit{in-situ} measurements.

\begin{table}[!htbp]
\caption{Summary of the results for all three events. The table lists the propagation angle, the arrival time, the mean velocity, and the impact velocity of the ICME for the used methods.}
\label{tab:results}
\begin{tabular}{lcccc}
\hline\noalign{\smallskip}
   02\,--\,06 June 2008      &       F$\phi$      &       HM       &      \textit{in-situ}  \\
\hline\\
propagation angle        &            E$24 \pm 3^\circ$        &  E$51 \pm 6^\circ$   &         --          \\
arrival day              &                --                   &        --        &         06 Jun.\ 2008 \\
arrival time             &                --                   &        --        &            15:35 \\
mean velocity [km~s\textsuperscript{$-1$}]     &           $389 \pm 48$             &   $395 \pm 35$   &       -- \\
impact velocity [km~s\textsuperscript{$-1$}]   &           $432 \pm 61$             &   $422 \pm 44$   &    $400$  \\ 
%\hline\noalign{\smallskip}
%\multicolumn{5}{c}{12--18 February 2009} \\
\hline\noalign{\smallskip}
  12\,--\,18 February 2009  &       F$\phi$      &        HM        &       \textit{in-situ}  \\                        
\hline\\
propagation angle       &            E$36 \pm 4^\circ$          &  E$61 \pm 3^\circ$   &        --          \\
arrival day             &                --                 &        --        &   18 Feb.\ 2010 \\
arrival time            &                --                 &        --        &    10:00 \\
mean velocity [km~s\textsuperscript{$-1$}]     &       $310 \pm 29$          &   $325 \pm 28$   &       -- \\
impact velocity [km~s\textsuperscript{$-1$}]   &       $416 \pm 35$        &   $398 \pm 34$ &   $362$  \\
%\hline\noalign{\smallskip}
%\multicolumn{5}{c}{03--05 April 2010} \\
\hline\noalign{\smallskip}
  03\,--\,05 April 2010      &     F$\phi$    &        HM        &      \textit{in-situ}  \\
\hline\\
propagation angle        &            W$3 \pm 4^\circ$              &  E$25 \pm 10^\circ$   &          --          \\
arrival day              &                --                   &        --        &               05 Apr.\ 2010 \\
arrival time             &                --                   &        --        &               07:58 \\
mean velocity [km~s\textsuperscript{$-1$}]     &           $829 \pm 122$             &   $854 \pm 100$   &          -- \\
impact velocity [km~s\textsuperscript{$-1$}]   &           $825 \pm 129$             &   $813 \pm 106$   &     $720$  \\ \hline\\
\end{tabular}
\end{table}

\section{Discussion and Conclusions}

The purpose of our study was to introduce a new technique, pointed out by \cite{moe09apj, moe10}, to additionally use the constraints imposed by \textit{in-situ} measurements to derive ICME kinematics end-to-end from the Sun to 1~AU. This method uses single-spacecraft heliospheric-imager data together with single-spacecraft \textit{in-situ} data and assumes a constant propagation direction.

Kinematics covering measurements in the HI1 and HI2 FoV were analyzed for three well-observed CMEs on their way from Sun to 1~AU. Images from the \textit{Heliospheric Imagers} onboard the NASA STEREO mission were used to produce time-elongation plots (Jmaps) from which the ICME measurements were derived. The measured elongation angles were converted into radial distances (in units of solar radii) by using two methods --- Fixed-$\phi$ (F$\phi$) and Harmonic Mean (HM) --- which approximate the ICME front as a point (F$\phi$) or a circle (HM). By combining remote sensing with \textit{in-situ} measurements it was possible to calculate the propagation directions as well as the kinematics of the events within the geometrical assumptions of the conversion methods. We constrained the ICME direction in such a way that the ICME distance-time and velocity-time profiles are most consistent with \textit{in-situ} measurements of the arrival time and the velocity on arrival. In \citet{tem11} the velocity profiles of the same three events were compared to the ambient solar wind, modeled by the ENLIL 3D MHD model \citep{ods03}.

The ICME of 02\,--\,06 June 2008 was clearly identified in the Jmap. It was a slow event ($\approx 390$ km~s\textsuperscript{$-1$}) that was embedded in the solar wind. It propagated toward STEREO-B, where clear signatures of a magnetic flux rope could be identified \textit{in-situ}. The F$\phi$ method reveals a propagation direction toward STEREO-B ($\phi_{F\phi} = $ E$24 \pm 3^\circ$), while the HM method assumes a wider longitudinal extent which may be improbable for this ICME event ($\phi_{HM} = $ E$51 \pm 6^\circ$), because the extent in the ecliptic should be around 25 degrees \citep{woo10}.
The method of \cite{woo10}, where synthetic images of a tube-like CME are adjusted to fit the CME's appearance in the STEREO/HI images, yields a direction in longitude of E$34^\circ$, and various directions derived by \cite{moe09apj} range from E$24^\circ$ to E$45^\circ$, all consistent with the directions derived in this paper.

The event of 13\,--\,18 February 2009 was a slow event too ($\approx 320$ km~s\textsuperscript{$-1$}). Its parameters were measured \textit{in-situ} by STEREO-B but do not show a clear three-part structure as for the June 2008 event. This CME originated from the limb and seems to have a wider longitudinal extension in the ecliptic and therefore was difficult to investigate since it dimmed rather quickly in the Jmap because it left the TS early. The results of the used methods differ again by about 25 degrees ($\phi_{F\phi}= $E$36 \pm 4^\circ$, $\phi_{HM}= $E$61 \pm 3^\circ$) and agree with the directions derived by \citet{moe11} who found E$35^\circ$ (F$\phi$ fitting) and E$63^\circ$ (HM fitting), respectively. The kinematics of both conversion methods are consistent with the \textit{in-situ} measured velocity.

The event of 03\,--\,05 April 2010 is an outstanding event for this solar minimum. In contrast to the other two events it was relatively fast ($\approx 835$ km~s\textsuperscript{$-1$}). The derived propagation directions range from $\phi_{F\phi}=$ W$3 \pm 4^\circ$ to $\phi_{HM}=$ E$25 \pm 10^\circ$ from Earth. Here, \cite{moe10} found a direction of longitude W$0 \pm 5^\circ$ using various methods (triangulation, Fixed-$\phi$ fitting, and forward modeling), and \cite{woo11} find a direction of W$2^\circ$, which makes our Fixed-$\phi$ result more consistent with the others quoted in the literature.
The velocity profile shows a different evolution for both methods. While F$\phi$ results in a nearly constant velocity between $\approx 750$ and 850~km~s\textsuperscript{$-1$}, HM shows an acceleration up to $\approx 100$ R$_{\odot}$ followed by a slight deceleration. The event seems to be decelerated from the slower ambient solar wind, in which the ICME was embedded \citep[\textit{e.g.}][]{gop01}.

The interpretation of the heliospheric images is rather difficult. Both conversion methods (Fixed-$\phi$ and Harmonic Mean) are useful to analyze the kinematics of ICMEs on their way through the heliosphere. How appropriate it is to use one or the other clearly depends on the chosen event. The big advantage of these methods compared to fitting methods such as the Sheeley-Rouillard \citep{she99,rou08} or the Harmonic Mean fitting method \citep{lug10,moe11} is that a constant velocity is not assumed. On the other hand they are not useable for forecasting because remote observations over the whole distance and \textit{in-situ} measurements at 1~AU are used as input to constrain CME kinematics. To assess the validity of the technique, numerical simulations should be used \citep[\textit{e.g.}][]{ods09,lug11}, because there the distance-time and velocity-time functions of the ICME front and its shape are known. Another disadvantage of the method is that for the calculation of the propagation direction only the remotely sensed elongation value at arrival time at 1~AU is used and constrained with \textit{in-situ} data. Assuming constant direction, the resulting direction is then used to convert the whole track of the remotely sensed CME. Using only one spacecraft from one vantage point it is not possible to derive the direction of every point along the CME-track in contrast to the triangulation technique by \citep{liu10a}.

All this is needed as a basis to be able to better forecast the direction and arrival time of coronal mass ejections using empirical or numerical propagation models. It will be necessary to investigate a large number of events to reveal their different kinematics in order to be able to apply the most appropriate method. Furthermore, the geometrical limitations of the F$\phi$ (point shaped CME) and HM (circle shaped CME) methods should be adjusted to consider the different widths of the CMEs.

%%%%%%%%%%%%%%%%%%%%%%%%%%%%%%%%%%%%%%%%%%%%%%%%%%%%%%%%%%%%%%%%%%%%%%%%%%%
%% Appendix
%
% \appendix   

%%%%%%%%%%%%%%%%%%%%%%%%%%%%%%%%%%%%%%%%%%%%%%%%%%%%%%%%%%%%%%%%%%%%%%%%%%%
%% Acknowledgements
%
 \begin{acks}
T.R., C.M.\ and M.T.\ were supported by the Austrian Science Fund (FWF): [P20145-N16]. The presented work has received funding from the European Union Seventh Framework Programme (FP7/2007-2013) under grant agreement n$^\circ$ 263252 [COMESEP]. It is also supported by NASA grants NNX10AQ29G and NAS5-03131. We thank the STEREO SECCHI/IMPACT/PLASTIC teams for their open data policy. We thank Thomas Rotter for carefully reading the manuscript.
 \end{acks}

%%% %%%%%%%%%%%%%%%%%%%%%%%%%%%%%%%%%%%%%%%%%%%%%%%%%%%%%%%%%%%

%% Bibliography
%
% Using BibTeX
%
 \bibliographystyle{spr-mp-sola}
% %\bibliographystyle{spr-mp-sola-cnd} %% Alternative style: no title, no concluding page
% \bibliography{rollett_bib_resub2}  
%

\end{article}
\end{document}